\documentclass[pdflatex,sn-mathphys-num]{sn-jnl}

\usepackage{graphicx}%
\usepackage{multirow}%
\usepackage{amsmath,amssymb,amsfonts}%
\usepackage{amsthm}%
\usepackage{mathrsfs}%
\usepackage[title]{appendix}%
\usepackage{xcolor}%
\usepackage{textcomp}%
\usepackage{manyfoot}%
\usepackage{booktabs}%
\usepackage{algorithm}%
\usepackage{algorithmicx}%
\usepackage{algpseudocode}%
\usepackage{listings}%
\usepackage{array}

\theoremstyle{thmstyleone}%
\theoremstyle{thmstyletwo}%

\theoremstyle{thmstylethree}%

\begin{document}

\title[Article Title]{From Demand to System Co-Shaping: A Review of User Roles in Transportation Systems}

\author*[1]{\fnm{Fangting} \sur{Zhou}}\email{fangting@chalmers.se}

\author[1]{\fnm{Balázs} \sur{Kulcsár}}\email{kulcsar@chalmers.se}

\author[2]{\fnm{Jelena} \sur{Andrić}}\email{jelenaa@chalmers.se}

\affil*[1]{\orgdiv{Electrical Engineering}, \orgname{Chalmers University of Technology}, \orgaddress{\city{Gothenburg}, \postcode{41296}, \country{Sweden}}}

\affil[2]{\orgdiv{Architecture and Civil Engineering}, \orgname{Chalmers University of Technology}, \orgaddress{ \city{Gothenburg}, \postcode{41296}, \country{Sweden}}}


\abstract{Users are central to transportation systems, yet their roles are often simplified in transportation modeling and decision-making. Conventional approaches primarily represent users through demand-related inputs, such as trip flows, delivery requests, and charging loads. However, digital, electrified, and platform-based services create more direct user–system interactions, with users responding to prices, incentives, service availability, and information in ways that can influence operational and planning decisions. This review develops a concept-oriented, taxonomy-based synthesis of how user-related information is represented and used in transportation decision-making. Rather than organizing the literature primarily by transport domain, we distinguish three user roles: passive demand, operational agents, and system co-shapers. Drawing on studies across transportation electrification, urban logistics, shared mobility, multimodal transport, and Mobility-as-a-Service (MaaS), the review examines how user preferences, flexibility, willingness to pay, acceptance, participation, and feedback are incorporated into modeling and decision frameworks. The synthesis shows that substantial progress has been made in measuring and predicting user behavior, but user-related variables are still often used mainly for behavioral analysis rather than as actionable inputs to system decisions. Their integration into operational decision-making remains incomplete, and systematic feedback from operational user responses and outcomes to long-term planning is even more limited. Key challenges include closing the loop between operation and planning, modeling heterogeneous and strategic users, integrating optimization with user acceptance, and ensuring privacy, equity, and deployability. Together, these findings point toward a shift from user-behavior analysis to user-informed and user-driven transportation system design.}


\maketitle

\section{Introduction}

Transportation systems are shaped by continuous interactions between users and system infrastructure. Travelers, vehicle operators, logistics customers, shared mobility users, and electric vehicle users make decisions about when, where, and how to use transportation services. These decisions collectively influence traffic flow, service utilization, energy consumption, infrastructure performance, and environmental outcomes. At the same time, system conditions, such as congestion, price, travel time, service availability, accessibility, and charging opportunities, affect how users adapt their behavior. This mutual interaction has long been central to transportation research, particularly in studies of route choice, travel demand, traffic assignment, user equilibrium, and discrete choice modeling \citep{OrtuzarWillumsen2011}.

Traditionally, however, users have often been represented as demand inputs in transportation planning and optimization models. Across passenger transport, freight and urban logistics, transportation electrification, and shared mobility, users are commonly represented through trip flows, delivery requests, charging demand, or service requests that serve as inputs to forecasting, planning, simulation, and optimization \citep{RasouliTimmermans2014,Boysen2021,Rashid2024,Laporte2015}. These demand-oriented representations remain useful, but they provide only a partial view of user influence.

Recent transformations in transportation systems are challenging this conventional view. Transportation electrification, shared mobility, Mobility-as-a-Service (MaaS), app-based logistics, and digital mobility platforms have created more direct interactions between users and system operators. Users respond to prices, waiting time, reliability, digital recommendations, service availability, and incentives. Their decisions affect short-term operational outcomes, such as station occupancy, fleet utilization, electricity load, routing efficiency, and service profitability. For example, electric vehicle users differ in charging behavior and may respond to charging prices or smart charging conditions \citep{will2016,motoaki2017,lee2020}. MaaS users respond to bundle composition, subscription design, and financial savings \citep{ho2018,matyas2019,ho2021}. Logistics customers respond to delivery fees, time windows, delivery speed, pickup-point accessibility, and parcel-locker distance \citep{nguyen2019,molin2022}. These studies suggest that users are not only sources of demand; they are operational agents whose choices and responses directly shape system performance.

Users can also influence long-term system planning. This role is particularly evident in transportation electrification, where charging behavior, location preferences, station utilization, service expectations, and technology adoption can inform charging station siting, capacity expansion, and service evolution \citep{li2018,babic2022,borlaug2023}. Similar patterns can be observed in shared mobility and mobility hub planning, where MaaS subscription behavior, hub design preferences, accessibility needs, and willingness to pay can influence service packages and infrastructure design \citep{bell2019,grigolon2025,silvestri2025,chen2026}. In urban logistics, customer choice and pickup-point acceptance can affect parcel-locker deployment, delivery network design, and self-collection infrastructure \citep{bruno2025,dayal2025}. These developments suggest that users increasingly contribute to transportation system evolution through behavioral feedback, service acceptance, and market interaction.

Although user behavior has been widely studied in transportation, existing research remains fragmented across application domains and decision levels. Relatively few studies provide a unified framework that captures the evolution of user roles from passive demand representations toward operational agents and system co-shapers across transportation domains.
Existing reviews often focus on specific domains, such as charging infrastructure and charging demand forecasting \citep{Farhadi2022,Rashid2024}, city logistics and last-mile delivery \citep{SavelsberghVanWoensel2016,Boysen2021}, MaaS \citep{Jittrapirom2017}, and shared mobility or bike-sharing systems \citep{Laporte2015}. A cross-domain synthesis of user roles and their implications for system-level decision-making is still needed. Table \ref{tab:positioning_existing_reviews} positions this review relative to existing user-oriented reviews and highlights its cross-domain decision-integration perspective.

\begin{table}[ht]
\centering
\caption{Positioning of this review relative to existing user-oriented transportation reviews.}
\label{tab:positioning_existing_reviews}
\begin{tabular}{@{}
    >{\raggedright\arraybackslash}
    p{\dimexpr0.20\linewidth-1.5\tabcolsep\relax}
    >{\raggedright\arraybackslash}
    p{\dimexpr0.24\linewidth-1.5\tabcolsep\relax}
    >{\raggedright\arraybackslash}
    p{\dimexpr0.26\linewidth-1.5\tabcolsep\relax}
    >{\raggedright\arraybackslash}
    p{\dimexpr0.30\linewidth-1.5\tabcolsep\relax}
@{}}
\toprule
\textbf{Review focus}
& \textbf{Representative reviews}
& \textbf{User perspective}
& \textbf{Gap addressed} \\
\midrule
Public transport mode choice and travel behavior
& \citet{suaa2022}; \citet{ranjan2025}
& Mode choice shaped by socio-demographic, trip, service, and contextual factors.
& Limited attention to operational and long-term planning influence beyond mode choice.
\\
Perceived accessibility and inclusive mobility
& \citet{kapsalis2024}; \citet{yarlagadda2025}
& Perceived accessibility, mobility barriers, inclusion needs, and travel experience.
& Limited translation into actionable inputs for pricing, scheduling, allocation, or infrastructure design.
\\
Integrated public transport and multimodal services
& \citet{chowdhury2016}; \citet{Jittrapirom2017}
& Responses to transfer quality, reliability, information, fare integration, and service bundles.
& Focus on service adoption and integration rather than user roles across decision levels.
\\
Shared mobility and micromobility
& \citet{Laporte2015}; \citet{badia2023}
& Adoption, trip purpose, use frequency, modal substitution, and service perceptions.
& Limited synthesis of users as operational agents or planning-feedback providers.
\\
Travel behavior change and intervention
& \citet{pan2025}
& Behavioral responses to incentives, feedback, nudges, and gamification.
& Limited integration of user responses and acceptance into operational optimization and planning.
\\
This review
& --
& Passive demand, operational agents, and system co-shapers.
& Cross-domain decision-integration lens linking user behavior with operations, planning, and system design.
\\
\botrule
\end{tabular}
\end{table}

To address this gap, this review makes three contributions.
First, it proposes a unifying decision-integration perspective on user roles in transportation systems. The literature is organized according to how user-related information is represented and used in system decisions: as passive demand, operational agents, and system co-shapers. This role-based taxonomy provides a cross-domain framework for understanding how users influence transportation systems.
Second, the review shows how user-related variables can be integrated into decision models, from demand estimation to pricing, scheduling, service allocation, infrastructure planning, and service redesign.
Third, it connects transportation electrification, urban logistics, shared mobility, multimodal transport, and MaaS through common user–system mechanisms, including flexibility, acceptance, willingness to pay, utilization feedback, and service rejection. This cross-domain synthesis further identifies practical requirements for deployable user-informed transportation systems, including data availability, operational KPIs, privacy, equity, interpretability, and industry implementation constraints.
Together, these contributions bridge the gap between academic user-behavior modeling and practical user-informed transportation system design.

The remainder of this paper is organized as follows. Section \ref{section2} defines the review scope and classification framework. Section \ref{section3} discusses studies that represent users as passive demand. Section \ref{section4} reviews research on users as operational agents. Section \ref{section5} examines users as system co-shapers in long-term planning. Section \ref{section6} compares modeling approaches across user roles. Section \ref{section7} discusses practical implementation, governance, and industry requirements. Section \ref{section8} proposes a research agenda for user-driven transportation system design. Section \ref{section9} concludes the paper.

\section{Review Scope and Decision-Integration Framework} \label{section2}

This review adopts a structured, concept-oriented, and taxonomy-building approach to examine how users are represented and integrated into transportation system decisions. The review synthesizes studies that explicitly connect user behavior, preferences, responses, participation, or feedback with system-level outcomes, including demand estimation, pricing, scheduling, service allocation, infrastructure utilization, facility location, capacity planning, and long-term system evolution.

In this review, user-driven transportation system design refers to the integration of user preferences, constraints, choices, participation, and feedback into transportation operation, planning, service design, and system adaptation. We use user-informed to describe systems in which user-related information supports system decisions, and user-driven to refer to stronger forms in which user inputs, responses, or feedback directly enter and shape operational or planning decisions.

\subsection{Review scope, search strategy and study selection}

The review scope is defined along three dimensions: transportation domain, user role, and system decision level. The application domains include transportation electrification, urban logistics, shared mobility, multimodal transport, and Mobility-as-a-Service (MaaS).

The review distinguishes three user roles, passive demand, operational agents, and system co-shapers, which correspond broadly to demand representation, operational decision-making, and long-term system planning, respectively.

A structured literature search was conducted using Scopus and Web of Science, supplemented by Google Scholar and backward and forward citation tracking. The primary search period covered studies published between 2010 and 2026, while earlier foundational contributions on travel demand modeling, user equilibrium, and discrete choice were retained to establish the theoretical background. Search terms combined user-related concepts such as “user behavior,” “user preference,” “willingness to pay,” “service acceptance,” and “user feedback” with system-level concepts related to transportation electrification, shared mobility, MaaS, multimodal transport, urban logistics, pricing, facility location, and transportation planning.

Studies were retained when they explicitly analyzed or modeled user-related factors and linked them to transportation system decisions or outcomes. Studies focusing solely on individual psychology, driving cognition, traffic safety behavior, vehicle control, battery technology, or supply-side optimization were excluded unless they established a direct connection with transportation system design, operation, or planning.

The selected studies were classified according to transportation domain, user role, and system decision level, with particular attention to how user-related information entered the decision process, as exogenous demand inputs, endogenous behavioral responses, or feedback signals for longer-term planning. This classification provided the basis for developing and refining the role-based taxonomy presented in Section \ref{section2.2}.

\subsection{Role-based taxonomy} \label{section2.2}

The resulting role-based taxonomy distinguishes three user roles according to how user-related information enters transportation decision models.

In passive-demand studies, users are represented primarily through exogenous parameters, observed states, or baseline constraints that support forecasting, assignment, resource allocation, and capacity planning but provide limited feedback to system decisions.

In operational-agent studies, users enter the decision process through endogenous choices, response functions, utility terms, acceptance constraints, participation decisions, or strategic interactions. Users may change their travel, charging, delivery, or service choices in response to system attributes, and these responses directly affect short-term outcomes such as congestion, occupancy, waiting time, routing efficiency, energy demand, and operator revenue.

In system co-shaping studies, user behavior, feedback, and associated operational outcomes are further translated into planning signals. Repeated congestion, low utilization, rejected offers, service non-use, location preferences, adoption patterns, and complaints may influence infrastructure location, capacity expansion, pricing structures, service coverage, and technology deployment. Explicit user participation through co-design, participatory planning, or living labs represents one form of co-shaping, but the concept also includes indirect influence through observed behavior and market feedback.

The three roles do not describe mutually exclusive user groups or a necessary progression followed by every user. Instead, they represent different ways of using user-related information in transportation decision-making. The same user may simultaneously generate demand, respond to operational conditions, and provide signals that influence future system design.

\subsection{Decision-integration framework}

\begin{figure}[ht]
\centering
\includegraphics[width=\textwidth]{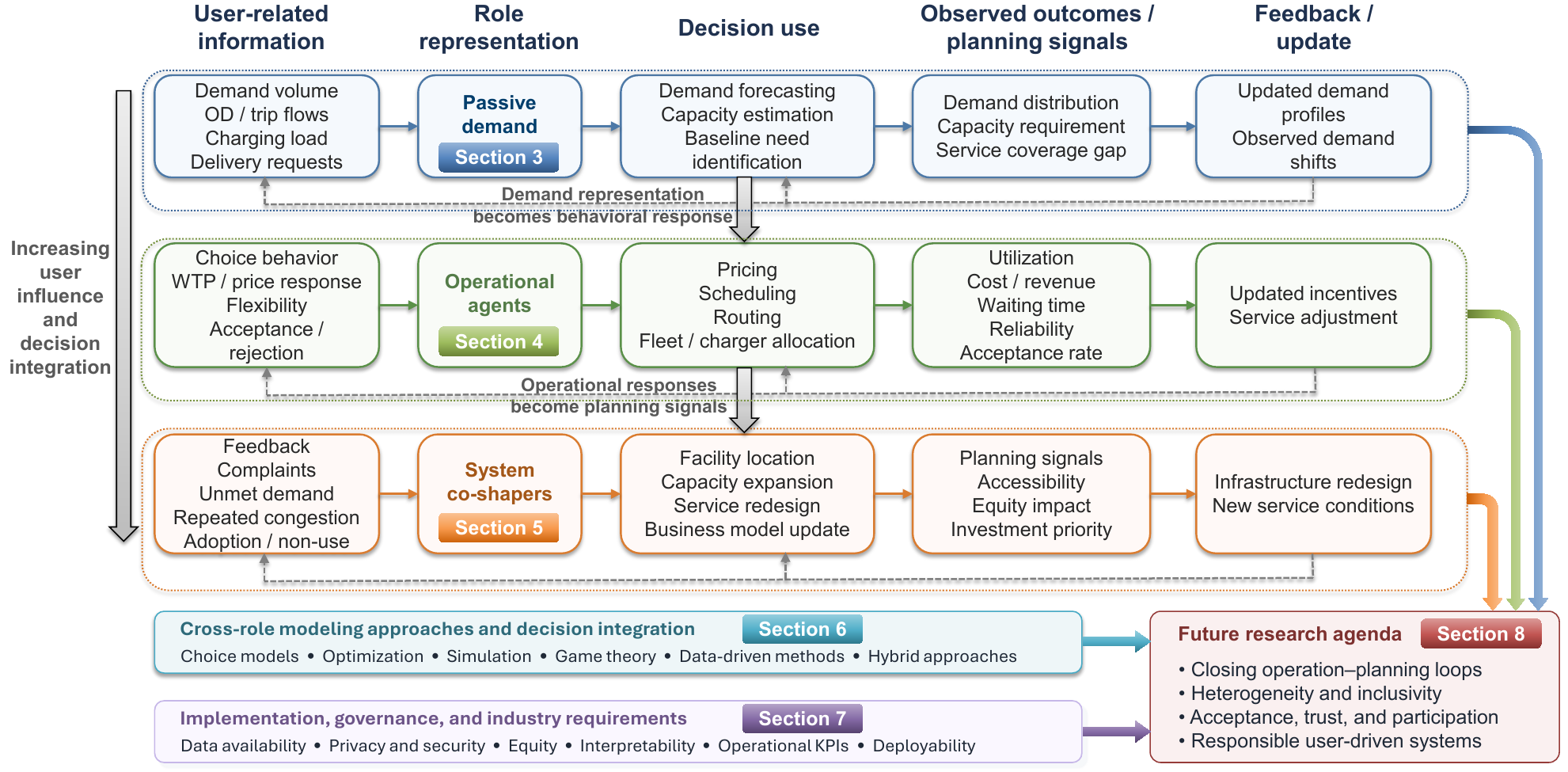}
\caption{Conceptual framework and paper roadmap for user-driven transportation system design.}\label{fig1}
\end{figure}

Figure \ref{fig1} presents the decision-integration framework used to organize the review. It connects user-related information, role representation, decision use, observed outcomes, and feedback to future decisions. The three horizontal layers correspond to passive demand, operational agents, and system co-shapers, while the vertical direction indicates increasing user influence and integration into transportation decision-making.

Across these layers, user-related information progresses from demand inputs to endogenous operational responses and, ultimately, to feedback signals for infrastructure planning and service redesign. Decisions implemented at each level subsequently alter the conditions experienced by users, creating feedback between users and transportation systems.
This framework structures the remainder of the review.

\section{Users as Passive Demand} \label{section3}

The most established representation of users in transportation research is as passive demand, where users are characterized primarily by the travel, charging, delivery, or service demand they generate. Behavioral mechanisms are typically simplified, aggregated, or treated as exogenous, providing tractable inputs for system decisions but limited representation of how users adapt to and influence transportation systems.

\subsection{Demand-based representation across transportation domains}

In passenger transportation, users are commonly represented through trip generation, origin–destination matrices, mode-specific demand, and network flows, which underpin conventional travel demand modeling, planning, and traffic assignment \citep{OrtuzarWillumsen2011}. Activity-based models provide a more disaggregate representation by linking travel to daily activities, schedules, and locations, although their outputs are still commonly translated into structured demand inputs for system-level planning and simulation \citep{RasouliTimmermans2014}.

In freight and urban logistics, customers are represented through order volumes, delivery locations, service time windows, and spatial demand distributions to support routing, facility location, and last-mile network design \citep{Boysen2021}. Users therefore influence the system primarily through the delivery requirements they generate rather than through responses to prices, locations, or service alternatives.

In transportation electrification, users are commonly represented through charging profiles, arrival and parking patterns, state of charge, required energy, and spatial charging demand. Such information supports charging-infrastructure planning and load management \citep{dong2014,Rashid2024}, while users remain primarily sources of charging demand to be forecasted or allocated.

Shared mobility and demand-responsive services use comparable representations, including trip requests, pickup–drop-off patterns, station-level demand, and lost demand, to support fleet planning, service coverage, and rebalancing \citep{Laporte2015}.

Across these domains, increasingly detailed observations may improve demand estimation without necessarily changing the passive role assigned to users in subsequent decision models.

\subsection{Limitations of passive-demand representation}

Passive-demand models remain essential for large-scale planning but have three main limitations. First, they often treat demand as fixed or only weakly responsive to system conditions, although users may change departure time, route, mode, charging location, delivery option, or service participation in response to price, waiting time, reliability, accessibility, and service availability. Second, aggregated demand can obscure heterogeneity in flexibility, willingness to pay, technology access, service expectations, and individual constraints. Third, passive-demand representations provide limited feedback to system design: they may indicate where and when demand occurs but not whether users will accept a proposed price, schedule, facility, or service configuration. These limitations become increasingly important in platform-mediated and demand-responsive systems and motivate representations in which users enter the decision process as active operational agents.

\section{Users as Operational Agents} \label{section4}

The second role in the proposed taxonomy views users as operational agents whose choices and responses directly influence short-term system performance. Rather than being represented only through the demand they generate, users respond to prices, time, reliability, accessibility, incentives, information, and service availability through decisions concerning travel, charging, service participation, and delivery options. These behavioral responses affect utilization, congestion, waiting time, routing and fleet efficiency, energy demand, service reliability, and operator revenue, providing a bridge between passive demand representation and longer-term system co-shaping.

\subsection{Behavioral responses to operational conditions}

Behavioral responses make realized demand endogenous to service conditions. EV users, for example, choose when, where, and how quickly to charge, with substantial heterogeneity in charging frequency, regularity, and flexibility \citep{dost2017,lee2020}. Charging-location decisions are shaped by accessibility, detour distance, waiting time, and surrounding activities, while prices, parking fees, and incentives can alter charging duration, station choice, and temporal demand \citep{Anderson2018,motoaki2017}.

The same mechanism appears in other mobility and logistics services. MaaS users respond to bundle composition, price, travel habits, and service attributes through subscription, booking, and mode-combination decisions; both stated-choice studies and real-world trials show that service uptake depends on service design and user characteristics \citep{ho2018,ho2021}. Last-mile customers similarly respond to delivery cost, time-window availability, convenience, and pickup-point accessibility when choosing among home delivery, service points, and parcel lockers \citep{nguyen2019,molin2022}. These choices directly influence realized demand and the operational conditions faced by service providers.

\subsection{Flexibility and operational participation}

Operational agency becomes stronger when users not only respond to service conditions but also provide flexibility that can be exploited in system operation. Smart charging, demand response, and vehicle-to-grid services provide a clear example: their technical benefits depend on users' willingness to shift charging or participate in coordinated charging. Optimization studies often treat such participation as given \citep{Farhadi2022,Shafiei2022}, whereas behavioral evidence shows that participation depends on compensation, perceived control, trust, charging reliability, and mobility requirements \citep{will2016}.

Similar forms of flexibility appear in other services. Incentives can encourage electric-carsharing users to accept alternative pickup or drop-off locations and thereby support vehicle relocation \citep{curtale2021}. In bike-sharing systems, users willing to walk to nearby stations when their preferred station is empty can reduce lost demand and support rebalancing decisions \citep{costa2021}. In urban logistics, customer flexibility between home delivery and shared locations can improve routing when appropriate compensation is provided \citep{mancini2021}, while passengers participating in public-transport-based crowdshipping become direct contributors to logistics operations, with participation shaped by compensation, additional time, and task characteristics \citep{fessler2022}. Across these cases, user flexibility becomes an operational resource, but its value depends on users' willingness to participate under the conditions offered.

\subsection{Integrating user responses into operational decisions}

A key challenge is to move from analyzing user responses to incorporating them directly into operational decisions. Behavioral studies can identify preferences, willingness to pay, acceptance, and participation, while optimization models often simplify these responses or treat them as exogenous assumptions. This separation is evident in flexible charging, where participation and behavioral constraints may determine whether technically optimal schedules can actually be implemented. More generally, predicting whether users will accept a price, service configuration, or operational request does not by itself ensure that such responses influence the decisions made by the operator.

Some studies increasingly bridge this gap by jointly representing user behavior and operational decisions. In EV charging, bid-informed frameworks can embed user-specified time windows, price bids, and energy requirements directly into operator pricing and scheduling decisions \citep{zhou2026}. In urban logistics, pricing and self-pickup incentives can be linked with customer responses and delivery operations \citep{islim2026}, while more integrated frameworks jointly model delivery alternatives, pricing, customer choice, and routing \citep{zhou2026last}. Such approaches shift user-related information from a descriptive or predictive input toward an endogenous component of operational decision-making. However, this integration remains uneven across transportation domains, and short-term user responses are rarely translated systematically into subsequent infrastructure or service-planning decisions. This limitation motivates the system co-shaping perspective examined in Section \ref{section5}.

\section{Users as System Co-Shapers} \label{section5}

The third role in the proposed taxonomy views users as system co-shapers. In this role, user behavior, preferences, participation, and feedback influence long-term infrastructure planning, service redesign, and system evolution. Unlike operational agents, whose responses affect immediate system performance, system co-shapers influence how repeated behavior, acceptance, non-use, utilization, complaints, and unmet demand are interpreted as signals for future planning and investment.

\subsection{From operational feedback to planning signals}

Operational interactions can generate planning signals when their effects persist or recur over time. Repeated charging congestion may indicate insufficient capacity; persistent rejection of service offers may reveal a mismatch between pricing, scheduling, and user constraints; high utilization may support facility expansion; and low adoption may indicate accessibility, pricing, or service-design problems. In this way, behavioral data, expressed preferences, utilization, acceptance, and broader market responses can inform facility location, capacity expansion, service redesign, investment prioritization, and technology deployment. The key shift is therefore from treating operational outcomes as end-point performance measures to using them as inputs for subsequent planning decisions.

\subsection{Translating user responses into infrastructure and service planning}

User-related information becomes co-shaping when it is translated into infrastructure or service-planning decisions. Charging infrastructure provides a clear example. User location preferences, detour tolerance, travel patterns, and charging demand can inform station siting and capacity decisions. Ouyang and Xu \cite{ouyang2022}, for example, incorporate detour tolerance and travel cost into charging-station siting, while Liagkas et al. \cite{liagkas2026} jointly optimize charger placement and dynamic spatio-temporal pricing while allowing price-sensitive users to shift charging location and time or opt out. Utilization and adoption provide additional feedback for infrastructure planning. EV adoption and charging infrastructure are mutually dependent, as insufficient infrastructure may discourage adoption while a limited user base may reduce investment incentives \citep{shi2021}. Local EV penetration and charging patterns are also important predictors of public charging-station utilization \citep{borlaug2023}, indicating how observed user behavior can inform future expansion and capacity decisions.

Similar mechanisms appear in shared mobility and mobility-hub planning. Repeated subscription behavior, willingness to pay, service uptake, and accessibility needs can inform service and infrastructure design, particularly when user responses differ across population groups \citep{silvestri2025,chen2026}. Mobility-hub studies further show how accessibility, transfer quality, walking distance, information provision, and other user requirements can shape hub design and location decisions \citep{bell2019,grigolon2025}. Accessibility-based planning can complement such evidence by translating commuting patterns, activity destinations, and existing public transport into mobility-hub location and service decisions \citep{frank2021}.

User responses can also influence the design of urban logistics networks. Choice-based facility-location studies show that willingness to use self-collection services can affect pickup-point and parcel-locker deployment \citep{bruno2025,dayal2025}. Customer flexibility, self-pickup incentives, and delivery-location choices can further affect routing performance and facility utilization, providing planning signals for delivery-network design and shared delivery infrastructure \citep{mancini2021,islim2026}. Across these applications, the central shift is from using user behavior only to describe demand to using behavioral and operational responses as inputs to infrastructure and service design.

\subsection{From implicit feedback to active co-shaping}

Observed behavior can provide valuable planning signals, but it should not automatically be interpreted as unconstrained user preference. Low use, for example, may reflect low demand, but it may also result from poor accessibility, limited information, exclusion, or unreliable service. Similarly, observed users represent only those able and willing to participate, while non-users and latent or constrained demand remain less visible. Translating operational feedback into planning decisions therefore requires distinguishing among preference, acceptance, access, and constraint.

A stronger form of co-shaping complements observed behavior with explicitly expressed user information. In charging services, for example, acceptable charging windows, minimum energy requirements, price limits, charging-speed preferences, and willingness to provide flexibility can reveal requirements that are difficult to infer from utilization alone \citep{zhou2026}. Acceptance and rejection of service offers can likewise indicate mismatches between user needs and existing pricing, capacity, or service conditions. Comparable signals can arise from repeated service use or non-use, willingness to pay, accessibility needs, and acceptance of alternative service configurations across mobility and logistics applications.

The distinction between implicit and active co-shaping therefore lies in whether user-related information is systematically interpreted and fed back into subsequent planning decisions. Evidence of such continuous feedback remains limited: most studies rely on stated preferences, observed usage data, or simulation-based planning, while relatively few demonstrate how repeated user responses lead to updated infrastructure or service decisions. This gap points to the need for stronger feedback loops between users, operations, and long-term planning.

\section{Modeling Approaches Across User Roles} \label{section6}

Different user roles imply different modeling requirements. Passive-demand models primarily estimate where, when, and how much demand occurs; operational-agent models capture endogenous choices and responses; and co-shaping approaches further connect user behavior and feedback with long-term planning and adaptation. Table \ref{tab:modeling_approaches} summarizes representative approaches, applications, strengths, and limitations across these roles.

\begin{table}[ht]
\centering
\caption{Modeling approaches across user roles in transportation systems.}
\label{tab:modeling_approaches}%
\begin{tabular*}{\textwidth}{@{\extracolsep\fill}
>{\raggedright\arraybackslash}p{0.12\textwidth}
>{\raggedright\arraybackslash}p{0.24\textwidth}
>{\raggedright\arraybackslash}p{0.11\textwidth}
>{\raggedright\arraybackslash}p{0.20\textwidth}
>{\raggedright\arraybackslash}p{0.18\textwidth}}
\toprule
\textbf{User role}
& \textbf{Representative approaches}
& \textbf{Typical applications}
& \textbf{Strengths}
& \textbf{Limitations} \\
\midrule
Passive demand 
& Travel demand forecasting; OD estimation; activity-based models; spatial demand modeling; statistical and machine learning prediction \citep{li2018,babic2022}
& Demand estimation and forecasting; capacity planning
& Scalable; compatible with large-scale planning, optimization, and simulation; useful for estimating demand distribution and resource needs
& Limited behavioral feedback; weak representation of heterogeneity, acceptance, participation, and system adaptation \\
Operational agents 
& Discrete choice models; latent class models; hybrid choice models; game theory; mechanism design; behavioral optimization; data-driven behavior inference \citep{ho2021,curtale2021,islim2026}
& Pricing, charging, service choice, routing, and rebalancing
& Captures user choices, preferences, incentives, price response, and heterogeneity; links service attributes with operational outcomes
& Often used for prediction rather than decision integration; may simplify strategic behavior, bounded rationality, and acceptance dynamics \\
System co-shapers 
& Facility location with user choice; accessibility-based planning; bilevel programming; agent-based simulation; digital twins; participatory planning; living labs \citep{ouyang2022,bruno2025,grigolon2025,aghaabbasi2025,liagkas2026}
& Facility location, capacity expansion, service design, and infrastructure redesign
& Connects user behavior with planning decisions; captures feedback, adaptation, participation, and long-term system evolution
& Data- and computation-intensive; difficult to validate; may face privacy, interpretability, governance, and implementation challenges \\
\botrule
\end{tabular*}
\end{table}

No single modeling approach fully captures user roles across these decision levels. Demand models provide scalable system inputs, behavioral models represent heterogeneous choices and responses, and feedback-oriented approaches connect user behavior with planning and adaptation. The methodological challenge is therefore to integrate these strengths without excessive behavioral, computational, or data requirements. Hybrid frameworks linking demand estimation, behavioral response, operational optimization, and planning feedback may provide a pathway toward decision-integrated transportation system design.

Decision integration also changes how transportation models should be evaluated. Operational efficiency alone is insufficient when system decisions depend on user participation and acceptance. System-side indicators such as cost, utilization, congestion, revenue, routing efficiency, and reliability should therefore be considered together with user-side outcomes such as acceptance, waiting time, willingness to pay, service rejection, and perceived fairness.
Distributional considerations such as equity may also determine whether an apparently efficient solution is practically viable. Computational tractability, scalability, and deployability are additional implementation considerations for data- and learning-intensive decision models \citep{zhou2025}.

Direct quantitative comparison across studies remains difficult because reported outcomes depend on context-specific service designs, behavioral assumptions, pricing mechanisms, and infrastructure constraints. Nevertheless, studies across EV charging, MaaS, and last-mile logistics show that user-related outcomes such as acceptance, willingness to pay, participation, and satisfaction can be evaluated alongside conventional operational indicators \citep{dai2021,tsouros2021,kokkinou2024,mohri2025,zhou2026}. Table \ref{tab:measurable_user_related_variables} therefore focuses on recurring user-related variables, their common measurement forms, decision relevance, and key interpretation cautions rather than proposing universal numerical benchmarks.

\begin{table}[ht]
\centering
\caption{Measurable user-related variables and their decision relevance.}
\label{tab:measurable_user_related_variables}%
\begin{tabular*}{\textwidth}{@{\extracolsep\fill}
>{\raggedright\arraybackslash}p{0.22\textwidth}
>{\raggedright\arraybackslash}p{0.22\textwidth}
>{\raggedright\arraybackslash}p{0.23\textwidth}
>{\raggedright\arraybackslash}p{0.18\textwidth}}
\toprule
\textbf{User-related variable}
& \textbf{Common measurement form}
& \textbf{Decision relevance}
& \textbf{Key caution} \\
\midrule
Price sensitivity/willingness to pay 
& Elasticity, stated or revealed WTP, bid, fare, or price response 
& Pricing, incentives, affordability assessment, demand management 
& Strongly context-, income-, and service-dependent \\
Time flexibility 
& Acceptable time window, delay tolerance, rescheduling willingness, accepted time shift 
& Charging scheduling, delivery windows, load shifting, time-slot allocation 
& Stated flexibility may differ from revealed behavior \\
Acceptance/rejection response 
& Acceptance rate, opt-in rate, offer response, service rejection 
& Feasibility of pricing, routing, service bundling, sharing, and flexible services
& Affected by trust, information, alternatives, and perceived control \\
Waiting and detour tolerance 
& Maximum waiting time, detour time or distance, queue tolerance 
& Facility location, service design, routing, pickup-point or locker design 
& Depends on trip purpose, urgency, and service reliability \\
Utilization and adoption feedback 
& Usage frequency, repeat use, non-use, complaints, unmet demand 
& Capacity expansion, infrastructure redesign, service adjustment, investment priority 
& Non-use may indicate lack of access rather than lack of demand \\
Perceived fairness/trust 
& Survey score, complaint rate, perceived markup, transparency rating 
& Governance, participation, platform design, policy support 
& Difficult to compare across studies, cultures, and service contexts \\
\botrule
\end{tabular*}
\end{table}

\section{Practical Implementation, Governance, and Industry Requirements} \label{section7}

Building on the role-based taxonomy and modeling comparison, this section examines the practical conditions for translating user-informed and user-driven models into deployable transportation systems, including regulatory compliance, market and platform governance, data availability, operational requirements, and industry usability.

\subsection{Regulatory, market, and societal considerations}

Regulatory considerations concern data protection, automated decision-making, and the use of personal information in transportation services. Mobility traces, app interactions, payment records, charging histories, and other user-related data may face restrictions on collection, reuse, storage, and sharing \citep{DeMontjoye2013,Primault2019}. Automated decisions involving personalized pricing, routing, charging allocation, or delivery-slot assignment may also raise concerns regarding transparency, explainability, consent where applicable, and mechanisms for user contestation. Regulation therefore affects both what user information can be used and how resulting decisions are implemented and communicated.

Market and platform conditions shape cross-provider coordination. User-informed services often depend on interoperability and data sharing, but these may be constrained by fragmented standards, limited willingness to open data access, and competing commercial interests \citep{Qiao2022}. Dynamic or personalized pricing may also face scrutiny when mechanisms are perceived as opaque, discriminatory, or anti-competitive. Platform governance and market structure therefore influence deployability alongside technical performance.

Societal considerations include acceptance, fairness, accessibility, trust, and public value. Efficiency-oriented solutions may fail when users perceive decisions as unfair, unreliable, difficult to understand, or insufficiently under their control. Pricing and allocation mechanisms may impose greater burdens on users with limited flexibility, while digital-only services may exclude users with limited digital access. Highly personalized services may also be less responsive to users with sparse behavioral data. User-informed transportation systems should therefore consider equity, accessibility, digital inclusion, perceived control, and trust alongside efficiency \citep{Martens2016,Pereira2017}.

Across these dimensions, important implementation gaps remain. Acceptance and trust are rarely represented directly in optimization models; transparency is often acknowledged without being translated into interpretable or auditable decision rules; privacy–performance trade-offs remain insufficiently quantified; and market and platform governance are seldom incorporated explicitly into transportation models. Addressing these gaps requires privacy-preserving and interpretable methods, fairness and acceptance criteria, and market-aware models that account for pricing, data sharing, interoperability, and platform coordination.

\subsection{Industry needs, practical requirements, and cautions}

Translating user-related information into deployable decision support requires behavioral variables that are observable, operationally relevant, and compatible with existing systems. In practice, this transition may begin with user-informed decision support, where user information improves planning and operational decisions before fully adaptive user-driven systems become feasible.

\subsubsection{Available user data and behavioral information gaps}

Current transportation systems generate extensive operational traces, including charging sessions, transactions, booking requests, vehicle telemetry, travel trajectories, app interactions, station occupancy, and customer-support records. These data reveal what users did and how services were used, but provide limited insight into why particular choices were made. Variables such as willingness to pay, flexibility, trust, perceived fairness, acceptance of automated decisions, and reasons for service rejection are rarely observed directly and often require surveys, stated-preference experiments, interviews, or active feedback mechanisms. The practical challenge is therefore to identify user variables that are both decision-relevant and feasible to collect, update, and maintain in operational systems.

\subsubsection{Passenger and commercial transportation requirements}

The requirements for user-informed transportation systems differ considerably between passenger and commercial applications. Passenger systems typically emphasize individual travel behavior, charging patterns, route choices, comfort, accessibility, and responses to prices or incentives. Commercial transportation places greater emphasis on fleet utilization, delivery schedules, driver hours, route compliance, loading patterns, and contractual service requirements, while user-related decisions must also be balanced against operational efficiency, regulatory requirements, and service-level commitments. The relevant decision-maker may therefore be a fleet manager, dispatcher, or logistics planner rather than the individual vehicle user. In shared charging systems, for example, fleet operators can act as the relevant infrastructure users and coordinate access to common charging resources under competing operational objectives \citep{zhou2026collaborative}. Commercial applications therefore involve multiple stakeholders whose objectives and constraints must be reconciled when designing user-informed decision support.

\subsubsection{Operational and user-centric performance metrics}

Transportation organizations typically evaluate performance using utilization, efficiency, reliability, cost, revenue, and service-quality indicators. Common examples include charger utilization and availability, delivery success and service-level compliance, and fleet utilization and vehicle availability. These operational KPIs remain essential, but user-informed systems also require indicators of behavioral feasibility and longer-term acceptance, such as offer acceptance, flexibility participation, service rejection, perceived fairness, customer effort, and retention. Performance frameworks should therefore combine operational and user-centric indicators to assess whether efficient services are also acceptable and sustainable from the user perspective.

\subsubsection{Practical considerations and cautions for deployment}

Practical deployment requires models that are interpretable, computationally tractable, compatible with existing operational platforms, and supported by appropriate data governance. Computational tractability is particularly important for large-scale operational optimization, where scalable solution methods can enable real-world scheduling applications \citep{mccabe2025}. Models may nevertheless have limited practical value if they depend on unavailable data or decision rules that operators and users cannot readily interpret. User-informed approaches therefore need to balance behavioral realism with operational usability.

Richer user data also do not automatically produce more user-centered decisions. Observed choices may reflect constraints in price, accessibility, information, availability, organizational rules, or viable alternatives rather than unconstrained preferences. Similarly, technical optimality does not guarantee behavioral feasibility: solutions that improve cost, utilization, congestion, or routing efficiency may still fail if they reduce convenience, reliability, fairness, trust, or perceived control. Successful deployment therefore requires operational efficiency to be considered alongside behavioral feasibility and broader objectives such as accessibility, equity, sustainability, and public value. Table \ref{tab:industry_requirements} summarizes these practical requirements by linking decision tasks with actionable user information, common data and interpretation challenges, and representative operational and behavioral KPIs.

\begin{table}[ht]
\centering
\caption{Practical requirements for user-informed transportation system design.}
\label{tab:industry_requirements}%
\begin{tabular*}{\textwidth}{@{\extracolsep\fill}
>{\raggedright\arraybackslash}p{0.17\textwidth}
>{\raggedright\arraybackslash}p{0.20\textwidth}
>{\raggedright\arraybackslash}p{0.26\textwidth}
>{\raggedright\arraybackslash}p{0.22\textwidth}}
\toprule
\textbf{Decision task}
&
\textbf{Potentially useful and observable user information}
&
\textbf{Common data gaps or interpretation challenges}
&
\textbf{Representative operational and behavioral KPIs}
\\
\midrule
Pricing (dynamic tariffs, incentives, service bundles)
& Transaction histories, charging records, booking behavior, app interactions, service-level selections
& True willingness to pay, perceived fairness, reasons for service rejection, long-term response to incentives
& Revenue, utilization, offer acceptance rate, conversion rate, customer retention \\
Scheduling and allocation (smart charging, fleet allocation, routing)
& Vehicle telemetry, charging duration, arrival/departure times, state-of-charge, route compliance, booking requests
& Distinguishing true flexibility from operational constraints, predicting last-minute changes, estimating user acceptance of proposed schedules
& Session completion rate, waiting time, flexibility participation, fleet utilization, routing efficiency, service reliability \\
Infrastructure planning (charging stations, hubs, depots, lockers)
& Usage records, occupancy patterns, geographic demand distribution, request volumes, fleet activity data
& Latent demand, non-user demand, future adoption behavior, long-term behavioral changes, interaction between physical and digital infrastructure
& Utilization, service coverage, capacity adequacy, return on investment, infrastructure availability \\
Service redesign (MaaS offerings, mobility services, delivery concepts)
& Customer-support interactions, repeated usage patterns, booking abandonment, service preferences, complaint records
& Root causes of churn, conflicting stakeholder preferences, fragmented data across operators, trust and acceptance factors
& Customer effort score, satisfaction, retention, adoption rate, delivery success rate, service-level compliance \\
Stakeholder-centered system design (passenger and commercial applications)
& Travel behavior, charging patterns, service preferences, fleet schedules, operational constraints, driver activity, contractual obligations, service requests
& Identifying who the relevant user is; reconciling competing objectives among travelers, drivers, fleet managers, operators, and customers; capturing organizational decision-making processes; transferring user-centered approaches across transportation domains
& Passenger: convenience, satisfaction, acceptance, retention. Commercial: fleet utilization, operational resilience, service-level compliance, regulatory compliance, contract performance \\
\botrule
\end{tabular*}
\end{table}

\section{Research Agenda: Toward User-Driven Transportation System Design} \label{section8}

Building on the role-based taxonomy, modeling comparison, and implementation requirements discussed above, this section outlines a research agenda for user-driven transportation system design. The central challenge is to translate user-related variables from descriptive or predictive measures into actionable inputs for operational decisions, planning signals, and adaptive system design. This requires frameworks that connect user responses with system outcomes across decision levels while accounting for behavioral heterogeneity, acceptance, privacy, equity, and deployability. Accordingly, the agenda focuses on six directions: behavior-to-decision integration, active user participation, closed-loop operation–planning feedback, heterogeneous and strategic users, optimization with acceptance, and adaptive, responsible, and privacy-preserving systems.

\subsection{From behavioral prediction to decision integration} \label{section8.1}

A first direction is to move from behavioral prediction to decision integration. Many studies can quantify user preferences, such as charging flexibility, willingness to pay, delivery preference, MaaS bundle preference, and service acceptance~\citep{will2016,ho2021,nguyen2019,molin2022,chen2026}. However, these variables are still often used mainly to explain or predict user choices.
Future research should further examine how these variables can be embedded into decision models as constraints, objectives, choice probabilities, acceptance thresholds, or feedback signals.

This shift is important because user-related variables can become decision inputs rather than only behavioral outputs. For example, acceptable charging time slots, minimum energy requirements, delivery flexibility, pickup-point acceptance, and MaaS subscription patterns can be embedded into charging scheduling, dynamic pricing, delivery routing, bundle design, and capacity expansion models. The key question is how to design models that are both behaviorally realistic and operationally usable.

\subsection{From implicit feedback to active user participation} \label{section8.2}

A second direction is to move from implicit behavioral feedback to active user participation. Current studies often infer preferences from observed service use, but such behavior may be ambiguous: non-use, for example, may reflect high price, poor accessibility, low reliability, limited information, lack of trust, or unavailable service rather than low demand.

Future systems should provide mechanisms through which users can explicitly express preferences, constraints, and willingness to participate. This is particularly important for services that depend on user cooperation, such as smart charging, vehicle-to-grid, self-pickup delivery, crowdshipping, and user-based relocation in shared mobility \citep{will2016,curtale2021,fessler2022,islim2026}. In such settings, user participation should be treated as a design input rather than only an observed outcome.

\subsection{Closing the loop between operation and planning} \label{section8.3}

A third direction is to develop closed-loop frameworks linking user responses, operational outcomes, and long-term planning. Existing planning models already consider utilization, demand, accessibility, and service coverage \citep{dong2014,frank2021,ouyang2022,bruno2025}, but user feedback is often represented indirectly or statically.

Future research should examine how repeated operational outcomes can become planning signals. Charging congestion may indicate a need for capacity expansion; low acceptance of smart-charging schedules may reveal inadequate compensation or reliability; and low parcel-locker adoption or MaaS renewal may indicate accessibility, trust, or service-design problems. User–system interaction should therefore be modeled as a continuous feedback process through which operational experience informs subsequent planning and service adaptation.

\subsection{Modeling heterogeneous and strategic users} \label{section8.4}

A fourth direction is to better represent user heterogeneity and strategic behavior. Users differ in value of time, price sensitivity, flexibility, accessibility constraints, charging access, digital access, trust, and service expectations, and such differences can materially affect system outcomes \citep{dost2017,ho2018,lee2020,silvestri2025,dayal2025}.

Future models should avoid relying only on representative users or average preferences when heterogeneity affects system outcomes. User segmentation, mixed logit models, latent class models, agent-based models, and data-driven clustering can support more differentiated service design. Strategic behavior also deserves greater attention, particularly in systems with dynamic pricing, priority rules, incentives, or compensation, where users may adjust their stated flexibility, urgency, or willingness to participate. Incentive-compatible and fairness-aware mechanisms are therefore needed to account for such responses.

\subsection{Integrating optimization with user acceptance} \label{section8.5}

A fifth direction is to integrate technical optimization with user acceptance. Many optimization studies focus on cost, travel time, routing efficiency, grid load, or infrastructure utilization, while participation and acceptance are simplified or assumed. Yet the reviewed evidence shows that technically attractive solutions may fail if users perceive them as inconvenient, unreliable, unfair, or insufficiently beneficial~\citep{will2016,polydoropoulou2022,karli2024,amaya2025,grigolon2025}.

Future optimization models should therefore treat acceptance as part of system feasibility and performance rather than as an exogenous assumption. Acceptance-related constraints, objectives, response functions, or participation probabilities can be used to connect technical decisions with user requirements concerning convenience, reliability, affordability, trust, and perceived control. Behavioral feasibility should thus be evaluated jointly with technical optimality.

\subsection{Toward adaptive, responsible, and privacy-preserving user-driven systems} \label{section8.6}

A sixth direction is to develop user-driven transportation systems that are adaptive, responsible, and privacy-preserving. Such systems increasingly rely on detailed behavioral and operational data, including mobility traces, transactions, service interactions, and user feedback. While these data can support more responsive pricing, scheduling, planning, and service design, they also raise concerns about privacy, data ownership, consent, and trust \citep{DeMontjoye2013,Primault2019}. Future research should therefore integrate privacy protection and data governance into system design rather than treating them as external implementation requirements. Privacy-preserving analytics should be accompanied by transparent rules concerning what data are collected, how they are used, and how users can contest or modify data-driven decisions.

Responsible design must also account for equity and public value. Data-driven systems may disproportionately respond to high-demand, high-income, or digitally active users while overlooking users with limited digital access, lower willingness to pay, or constrained mobility options. User-driven design should therefore not become purely market-driven design; accessibility, distributional effects, and fairness should be evaluated alongside efficiency and responsiveness \citep{Martens2016,Pereira2017}.

These concerns become increasingly important as AI-mediated platforms and recommendation systems shape information access and user choices. Users may base travel and service decisions on system-wide information about travel time, prices, charging availability, emissions, reliability, and congestion, potentially improving individual decision quality and system responsiveness \citep{arora2026}. 
However, AI-mediated recommendations may also reshape demand, concentrate or shift congestion, and steer users toward choices that are individually rational but collectively undesirable \citep{cornacchia2026}. Unequal access to information may introduce additional distributional concerns.
Future research should therefore examine how platform-mediated information affects demand formation and system externalities, and how users, platforms, and public authorities should share responsibility for these outcomes.

Table~\ref{tab:future_directions} summarizes the six research directions discussed in this section by linking each future direction with a core research question and a key caution.

\begin{table}[!t]
\centering
\small
\caption{Future research agenda and key cautions for user-driven transportation system design.}
\label{tab:future_directions}%
\begin{tabular*}{\textwidth}{@{\extracolsep\fill}
>{\raggedright\arraybackslash}p{0.22\textwidth}
>{\raggedright\arraybackslash}p{0.36\textwidth}
>{\raggedright\arraybackslash}p{0.32\textwidth}}
\toprule
\textbf{Future direction}
& \textbf{Core research question}
& \textbf{Key caution} \\
\midrule
Behavior-to-decision integration 
& How can quantified user preferences directly inform pricing, scheduling, allocation, and planning? 
& Avoid using user variables only for behavior prediction or explanation. \\
Active user participation 
& How can users explicitly express preferences, flexibility, willingness to pay, and acceptance? 
& Avoid treating users only as passive data sources. \\
Closed-loop operation--planning feedback 
& How can operational user feedback inform long-term infrastructure and service planning? 
& Avoid treating user feedback as static demand rather than dynamic planning signals. \\
Heterogeneous and strategic users 
& How do different user groups respond to incentives, prices, and service designs? 
& Avoid relying only on representative users or average preferences. \\
Optimization with acceptance 
& How can technical optimization account for trust, control, reliability, and participation? 
& Avoid assuming users automatically accept technically optimal solutions. \\
Adaptive, responsible, and privacy-preserving systems 
& How can user-driven systems remain adaptive, privacy-preserving, equitable, and responsible under data-intensive platforms and AI-mediated decisions?
& Avoid purely data- or market-driven design; privacy, consent, equity, and public value must remain central as AI-mediated choices reshape demand and externalities. \\
\botrule
\end{tabular*}
\end{table}

\section{Conclusion} \label{section9}

Users have always been central to transportation systems, but their role has often been simplified in transportation modeling and decision-making. This review argues that the role of users is evolving from passive demand representation toward active participation in system operation and long-term system design. This evolution is particularly visible in emerging transportation systems, such as transportation electrification, urban logistics, shared mobility, multimodal transport, and MaaS, where users interact with digital platforms, infrastructure, pricing mechanisms, and service providers more directly than before.

Based on this perspective, this review proposes a role-based taxonomy of users in transportation systems. The first role represents users as passive demand, where users are mainly described by the amount, location, timing, or type of demand they generate. The second role views users as operational agents, whose choices and responses influence short-term system performance, including pricing, scheduling, routing, charging, fleet operation, and service utilization. The third role views users as system co-shapers, whose behavior, preferences, participation, and feedback can inform infrastructure planning, capacity expansion, service design, technology deployment, and system evolution.

The reviewed literature shows that transportation research has made substantial progress in measuring and modeling user behavior. User preferences, such as time flexibility, willingness to pay, detour tolerance, service acceptance, charging requirements, and delivery preferences, can increasingly be quantified through surveys, discrete choice models, behavioral data, platform records, and data-driven methods. However, these variables are still often used mainly for demand prediction or behavioral explanation. A key future challenge is to translate user-related variables into decision-relevant inputs and behavioral KPIs that can support pricing, scheduling, resource allocation, infrastructure planning, and long-term system adaptation.

Future transportation systems will require stronger feedback between users and system design. This calls for models that integrate user participation, operational decisions, and long-term planning; account for heterogeneity and strategic behavior; combine technical optimization with user acceptance; and address equity, privacy, and responsible data use. As AI-mediated platforms and recommendation systems increasingly shape user decisions, user-driven design should also consider how information access, platform incentives, and individual choices influence demand formation and system externalities.
Deployable user-driven systems must also address data governance, interoperability, regulatory compliance, operational KPIs, and industry usability.
Moving from user behavior analysis to user-driven transportation system design can help build systems that are not only efficient but also adaptive, inclusive, responsible, and responsive to the people they serve. In this sense, the proposed taxonomy provides not only a way to organize existing literature but also a conceptual foundation for designing transportation systems in which users are treated as active contributors to operational performance, infrastructure planning, and system evolution.

\section*{Declarations}

\bmhead{Funding}
This work was supported by the European Commission and the Swedish Energy Agency (Grant No. F-ENUAC-2022-0003), and the Swedish Electromobility Centre (Grant No. 11129).

\bmhead{Authors' contributions} Fangting Zhou: Conceptualization, formal analysis, funding acquisition, investigation, methodology, visualization, writing—original draft, and writing—review and editing. Balázs Kulcsár: Conceptualization, funding acquisition, methodology, supervision, visualization, and writing—review and editing. Jelena Andrić: Conceptualization, formal analysis, visualization, and writing—review and editing.

\bmhead{Competing interests}
The authors declare that they have no competing interests.

\bibliography{sn-bibliography}

\end{document}